\newcommand{\be}{\begin{equation}}\newcommand{\ee}{\end{equation}}
\newcommand{\bea}{\begin{eqnarray}}\newcommand{\eea}{\end{eqnarray}}
\newcommand{\nn}{\nonumber}\newcommand{\p}[1]{(\ref{#1})}
\documentstyle[12pt]{article}

\begin{document}
\renewcommand{\thefootnote}{\fnsymbol{footnote}}
\thispagestyle{empty}
\begin{center}
{\hfill LNF-96/041 (P)}\vspace{0.2cm} \\
{\hfill hep-th/9609038}\vspace{2cm} \\
{\bf The $W(sl(N+3),sl(3))$ algebras and their contractions 
to $W_3$}\footnote{Dedicated to the
memory of Victor I. Ogievetsky, as the expression of our respectful
admiration and deep sorrow.}
\vspace{1.5cm} \\
S. Bellucci${}^a$\footnote{E-mail: bellucci@lnf.infn.it} and
S. Krivonos${}^{b}$\footnote{E-mail: krivonos@thsun1.jinr.dubna.su} 
and 
A. Sorin${}^{b}$\footnote{E-mail: sorin@thsun1.jinr.dubna.su} 
                    \vspace{1cm} \\
${}^a$INFN-Laboratori Nazionali di Frascati, P.O.Box 13 I-00044 Frascati,
            Italy \\
${}^b$Bogoliubov Laboratory of Theoretical Physics, JINR, Dubna, Russia
\vspace{2.5cm} \\
{\bf Abstract} 
\end{center}
We construct the nonlinear
$W(sl(N+3),sl(3))$ algebras and
find the spectrum of values of the central charge
that gives rise, by contracting the
$W(sl(N+3),sl(3))$ algebras,
to a $W_3$ algebra belonging to the
coset $W((sl(N+3),sl(3))/(u(1)\oplus sl(N))$.
Part of the spectrum was conjectured before, but part of it is given here
for the first time. Using the tool of embedding the
$W(sl(N+3),sl(3))$ algebras into linearizing algebras, we construct
new realizations of $W_3$ modulo null fields.
The possibility to predict, within the
conformal linearization framework, the central charge
spectrum for minimal models of the
nonlinear $W(sl(N+3),sl(3))$ algebras is discussed at the end.
\vspace{1.5cm} \\
\vfill
\begin{center}
September 1996
\end{center}
\setcounter{page}0
\renewcommand{\thefootnote}{\arabic{footnote}}
\setcounter{footnote}0
\newpage

\section{Introduction}
                                      
$W$-algebras were introduced in 1985 by Zamolodchikov \cite{z}, in
terms of conformal models, exhibiting new symmetries generated by
currents with conformal spin higher than 2. 
Owing to the nonlinearity of $W$-algebras, the important
task of constructing their realizations in terms of free fields or affine
currents cannot be carried out straightforwardly. 

One of the possible ways to construct the realizations of $W$ algebras
is the conformal linearization procedure \cite{a4,a5,ks1,a13,a6}. 
The main idea of this
approach is to embed the nonlinear $W$ algebra as a subalgebra
in some linear conformal algebra $W^{lin}$. Once this is done, 
then each realization 
of the linear algebra $W^{lin}$ gives rise to a realization of $W$.

In \cite{a6} we constructed explicitly the nonlinear algebras
$W(sl(4),sl(3))$, $W(sl(3|1), sl(3))$ and obtained their realizations
in terms of currents spanning the corresponding
linearizing conformal algebras.
The specific structure of these algebras allowed us to
construct realizations
- modulo null fields - of the $W_3$ algebra that lies in 
the cosets $W(sl(4),sl(3))/u(1)$ and $W(sl(3|1),sl(3))/u(1)$. In
such null-fields realizations the OPE of two spin-3 currents
contains a spin-4 operator which is {\it null}, in the sense that
there is no central term in the OPE of two such spin-4 currents.
This occurs only for a discrete spectrum of
$W(sl(4),sl(3))$ - respectively $W(sl(3|1), sl(3))$ - central charges,
which allows us to consider a vanishing value for such spin-4 operators
and reduces the $W(sl(4),sl(3))$ - respectively $W(sl(3|1), sl(3))$ -
algebra to its $W_3$ contraction. For this spectrum the realizations
of $W(sl(4),sl(3))$ - respectively $W(sl(3|1), sl(3))$ - algebras
provide null-fields realizations of $W_3$.

In this Letter we follow a similar procedure, in order to
obtain new null-fields realization of $W_3$ from linearizing
$W((sl(N+3),sl(3))$ algebras. In particular, we
determine the central charge spectrum for the
nonlinear $W(sl(N+3),sl(3))$ algebras, when these algebras are
contracted to the $W_3$ algebra lying in the
coset $W((sl(N+3),sl(3))/(u(1)\oplus sl(N))$.
The completion of our task, namely the explicit construction
of free-fields realizations - modulo null fields - for the $W_3$
contraction of $W(sl(N+3),sl(3))$ algebras, becomes straightforward
when the conformal linearization procedure is applied to the latter
algebras. Indeed, it turns out that the nonlinear
$W(sl(N+3),sl(3))$ algebras can be embedded into some linear
conformal algebras with a finite number of generators
related to the currents of the nonlinear algebra by
an {\it invertible} transformation. Hence it follows that
any realization of the linearizing
$W(sl(N+3),sl(3))$ algebras yields a realization of the nonlinear
$W(sl(N+3),sl(3))$ ones.

The outline of the Letter is as follows.

We start in Section 2 with presenting
the nonlinear $W(sl(N+3),sl(3))$ algebras. Their explicit structure appears
here for the first time.
The importance of the algebras constructed in this Section can be
appreciated if one recalls that further examples of analogous algebras
whose OPEs explicitly depend, in addition to the level $K$ of the algebra,
on one more parameter (in our case denoted by $N$) are very few.
They include the Knizhnik-Bershadsky algebras, as well as the
quasi superconformal algebras.
In Section 3, after recalling briefly the method used to
construct null-fields realizations of $W_3$, we
consider the exceptional values of the central charges of
the nonlinear $W(sl(N+3),sl(3))$ algebras 
for which the composite spin-4 field
appearing in the spin-3 - spin-3 OPE becomes a null field.
Also we discuss the properties of the spectrum and point out
the result that one central charge value corresponds to three
distinct representations of the contracted algebra $W_3$.
We then proceed in Section 4 to construct
the corresponding linear conformal algebras with a finite number of
currents. The resulting expressions giving,
in terms these currents, those 
spanning the nonlinear algebras immediately provide
the realizations of $W(sl(N+3),sl(3))$.
The constructed expressions for the currents of the nonlinear algebras
in terms of the linearizing algebras currents yield, for the  specific 
values of the central charge, the realization
of $W_3$ algebra modulo null fields.

We will finish this Letter with some concluding remarks and a short
discussion of further developments, including applications
to the central charge spectrum for minimal models of
the nonlinear $W(sl(N+3),sl(3))$ algebras.

\section{Nonlinear $W(sl(N+3),sl(3))$ algebras}

In this Section we give explicitly for the first time the structure 
of the nonlinear $W(sl(N+3),sl(3))$ algebras in the quantum case.
This result provides a new example
of algebras that depend on two parameters, i.e. the
level $K$ and $N$, and possess the $SL(N)$ automorphism. The constructed
$W(sl(N+3),sl(3))$ algebras
are analogous in this sense to the case of the Knizhnik-Bershadsky
and quasi superconformal algebras.

Let us first of all describe the conformal spin content
of these algebras \cite{a2}. They are spanned, in addition to the spin-2
stress-tensor
$T$, by the following currents: the $sl(N)$ and $u(1)$
affine spin-1 currents $J_a^b$ and $U$, the commuting-with-them spin-3
current $W$ and two multiplets of spin-2 currents $G_a$,
${\overline G}^a$ with opposite $u(1)$ charges, which belong to the
fundamental and its conjugated representations of $sl(N)$.
In the basis where these bosonic currents
$(U,J_a^b,G_a,{\overline G}^a,W)$ are all
primary fields with respect to $T$, the singular OPEs
of the nonlinear $W(sl(N+3),sl(3))$ algebras read
\begin{eqnarray}
T(z_1)T(z_2) & = &  \frac{c_t}{2z_{12}^4}+
        \frac{2T}{z_{12}^2}+\frac{T'}{z_{12}}\quad ,
\quad
T(z_1)J_a^b (z_2) = \frac{J_a^b}{z_{12}^2}+\frac{J_a^b{}'}{z_{12}}
\quad ,\nn\\
T(z_1)U(z_2) & = &  \frac{U}{z_{12}^2}+\frac{U'}{z_{12}} \quad ,
\quad
T(z_1)W(z_2) = \frac{3W}{z_{12}^2}+\frac{W'}{z_{12}} \quad ,\nn  \\
T(z_1)G_a(z_2) & = &  \frac{2G_a}{z_{12}^2}+\frac{G_a'}{z_{12}}\quad , 
\quad
T(z_1){\overline G}^a(z_2) = \frac{2{\overline G}^a}{z_{12}^2}+
\frac{{\overline G}^a{}'}{z_{12}} \quad ,\nn \\ 
U(z_1)U(z_2) & = &  \frac{c_u}{z_{12}^2} \quad , 
\quad
U(z_1)G_a(z_2) =  -\frac{2G_a}{z_{12}} \quad ,
\quad
U(z_1){\overline G}^a(z_2) = \frac{2{\overline G}^a}{z_{12}} \quad ,\nn  \\
J_a^b (z_1)J_c^d(z_2) & = & K(\delta_a^d\delta_c^b -\frac{1}{N}
\delta_a^b\delta_c^d )\frac{1}{z_{12}^2}+(\delta_c^b J_a^d -
\delta_a^d J_c^b )\frac{1}{z_{12}}\quad ,\nn \\
J_a^b (z_1)G_c(z_2) & = & (\delta_c^b G_a -\frac{1}{N}
\delta_a^b G_c )\frac{1}{z_{12}} \quad , 
\quad
J_a^b (z_1){\overline G}^c(z_2) =  (-\delta_a^c {\overline G}^b
+\frac{1}{N}\delta_a^b {\overline G}^c )\frac{1}{z_{12}} \quad ,  \nn  \\
G_a(z_1){\overline G}^b(z_2) & = &  \delta_a^b\frac{c_g}{z_{12}^4}+
   ({\alpha}_1 \delta_a^b U+{\alpha}_{2}J_a^b)\frac{1}{z_{12}^3}+
   \left[ \frac{1}{3} \delta_a^b T+{\alpha}_3 \delta_a^b (U\;U)+
   \frac{\alpha_1}{2} \delta_a^b U' 
 +  {\alpha}_4 (J_a^c\;J_c^b) 
\right. \nn \\
     & & \left.
  + \frac{\alpha_1}{K} (U\;J_a^b)
  + {\alpha}_5 \delta_a^b (J_c^d\;J_d^c)
  + K{\alpha}_4 J_a^b {}'  \right]
   \frac{1}{z_{12}^2}+
   \left[ \delta_a^b W+\frac{1}{6} \delta_a^b T'+
   {\alpha}_{6} \delta_a^b (T\;U)
\right. \nn \\
     & & \left.
+{\alpha}_{7}\delta_a^b (U\;(U\;U))
+   {\alpha}_{3}\delta_a^b (U'\;U)
+   {\alpha}_{8}\delta_a^b U''
+   {\alpha}_{9}\delta_a^b (U\;(J_c^d\;J_d^c))
\right. \nn \\
     & & \left.
+   {\alpha}_{10}\delta_a^b (J_c^d\;(J_d^e\;J_e^c))
+   {\alpha}_{11}\delta_a^b (J_c^d {}'\;J_d^c)
+   \frac{1}{3K} (T\;J_a^b)
+   \frac{\alpha_3}{K} (U\;(U\;J_a^b))
\right. \nn \\
     & & \left.
+   {\alpha}_{12} (U\;(J_a^c\;J_c^b))
+   {\alpha}_{13} (J_a^b\;(J_c^d\;J_d^c))
+   2{\alpha}_{13} (J_a^c\;(J_c^d\;J_d^b))
+   \frac{\alpha_1}{2K} (U'\;J_a^b)
\right. \nn \\
     & & \left.
+   K{\alpha}_{12} (U\;J_a^b {}')        
+   {\alpha}_{14} (J_a^c {}'\;J_c^b)
+   2K{\alpha}_{13} (J_a^c\;J_c^b {}')
+   {\alpha}_{15} J_a^b{}''  \right]
   \frac{1}{z_{12}}  \quad ,   \nn \\
G_a(z_1)W(z_2) & = &  \frac{{\beta}_1 G_a}{z_{12}^3}+
         \left[ {\beta}_2 G_a{}'+{\beta}_3 (U\;G_a)
        +{\beta}_4 (J_a^b\;G_b) \right]
         \frac{1}{z_{12}^2} \nn \\
   & & +\left[ {\beta}_5 (T\;G_a)+{\beta}_6 (U\;(U\;G_a))
  +{\beta}_7 (U\;(J_a^b\;G_b))
  +{\beta}_8 (U'\;G_a)+{\beta}_9 (U\;G_a{}') \right. \nn \\
     & & \left.
  +{\beta}_{10} (J_a^b\;(J_b^c\;G_c))
  +{\beta}_{11} (J_b^c\;(J_c^b\;G_a))
  +{\beta}_{12} (J_a^b{}'\;G_b) +{\beta}_{13} (J_a^b\;G_b{}') 
\right. \nn \\
     & & \left.
  +{\beta}_{14} G_a{}''\right]
    \frac{1}{z_{12}}  \quad , \nn  \\
{\overline G}^a(z_1)W(z_2) & = & 
  \frac{{\beta}_1 {\overline G}^a}{z_{12}^3}+
  \left[ {\beta}_2 {\overline G}^a{}'-{\beta}_3 (U\;{\overline G}^a)
  -{\beta}_4 (J_b^a\;{\overline G}^b)
  \right] \frac{1}{z_{12}^2}\nn\\
     & &
+     \left[ {\beta}_5 (T\;{\overline G}^a)+{\beta}_6
    (U\;(U\;{\overline G}^a))+{\beta}_7 (U\;(J_b^a\;{\overline G}^b))
  -{\beta}_8 (U'\;{\overline G}^a{}')-{\beta}_9 (U\; {\overline G}^a{}')
\right. \nn \\
    & & \left.
  +{\beta}_{10} (J_b^c\;(J_c^a\;{\overline G}^b))
  +{\beta}_{11} (J_b^c\;(J_c^b\;{\overline G}^a))
  -{\beta}_{12} (J_a^b{}'\;{\overline G}^b))
  -{\beta}_{13} (J_a^b\;{\overline G}^b{}')
\right. \nn \\
     & & \left.
  +{\beta}_{14} {\overline G}^a{}''\right]
     \frac{1}{z_{12}}    \quad , \nn \\
W(z_1)W(z_2) & = &  {\epsilon} \left\{
  \frac{c_w}{6z_{12}^6}+
      T_w
     \frac{1}{z_{12}^4}+\frac{1}{2}
     T_w '
       \frac{1}{z_{12}^3}\right. \nn \\
   & &
+  \left.   \left[ U_4 +
\frac{16}{22+5c_w}
\left( (T_w \;T_w )-\frac{3}{10}T_w ''\right)
+\frac{3}{20}T_w ''
    \right]
   \frac{1}{z_{12}^2}\right. \nn \\
    & &
+\left. \frac{1}{2} \left[ U_4 +
\frac{16}{22+5c_w}
\left( (T_w \;T_w )-\frac{3}{10}T_w ''\right)
+\frac{1}{15}T_w ''
    \right] '
   \frac{1}{z_{12}} \right\} \label{superWnew} ,
\end{eqnarray}
where 
\begin{eqnarray}
T_w & = & T-\frac{(U\;U)}{2c_u}-\frac{(J_a^b\;J_b^a)}{2(K+N)}
\quad , \nn  \\
c_{w} & = & -\frac{(2K+N)(-1+3K+2N)(1+4K+3N)}{(K+N)(K+N+1)}
\quad , \nn  \\
U_4 & = & {\epsilon}_1W'+{\epsilon}_{2}(W\;U)+{\epsilon}_{3}(G_a\;
{\overline G}^a)+{\epsilon}_{4}(T\;T)+{\epsilon}_{5}(T\;(U\;U))
+{\epsilon}_{6}(T\;(J_a^b\;J_b^a))
    \nn \\
    & & 
+{\epsilon}_{7}(T'\;U)
+{\epsilon}_{8}
(T\;U')+{\epsilon}_{9}T''
+{\epsilon}_{10}(U\;(U\;(U\;U)))+{\epsilon}_{11}(U'\;(U\;U))
\nn \\
    & &
+{\epsilon}_{12}(U\;(U\;(J_a^b\;J_b^a)))
+{\epsilon}_{13}(U''\;U)
+{\epsilon}_{14}(U'\;U')+{\epsilon}_{15}U'''
+{\epsilon}_{16}(U\;(J_a^b\;(J_b^c\;J_c^a)))
\nn \\
    & &
+{\epsilon}_{17}(U'\;(J_a^b\;J_b^a))
+{\epsilon}_{18}(U\;(J_a^b{}'\;J_b^a))+{\epsilon}_{19}
(J_a^b\;(J_b^c\;(J_c^d\;J_d^a)))+{\epsilon}_{20}
(J_a^b\;(J_b^a\;(J_c^d\;J_d^c)))
\nn \\
    & &
+{\epsilon}_{21}(J_a^b{}'\;(J_b^c\;J_c^a))
+{\epsilon}_{22}(J_a^b{}'\;J_b^a{}')
+{\epsilon}_{23}(J_a^b{}''\;J_b^a)
\nn \\
    & &
-\frac{16}{22+5c_w}
\left[ (T_w \;T_w )-\frac{3}{10}T_w ''\right]
-\frac{3}{20}T_w ''
\quad . \label{ucoset} 
\end{eqnarray}
Here the indices $a,b,...$ run over the following ranges
$1\leq a,b \leq N$, and the $c_{t,u,g}$, $\alpha$, 
$\beta$ and $\epsilon$ coefficents are found in Table 1 (see appendix)
in terms of the level $K$ of $sl(N)$ algebra.
Although the coefficient ${\epsilon}$ could be set to unity
provided the current $W(z)$ is rescaled, we keep it for convenience,
since this makes the coefficients simpler.

Inspecting the expressions given in Table 1 makes it evident
that the OPE $T(z_1)T(z_2)$ (\ref{superWnew}) is singular for $K=-1-N$. 
In fact,
the $W(sl(N+3),sl(3))$ algebras are not defined for this level
value\footnote{After rescaling their currents, the algebras do not contain
the
Virasoro subalgebra.} and we take $K\neq -1-N$. In order to remove from the 
OPEs (\ref{superWnew}) the additional singularities occurring in some
of the coefficients of Table 1 for the level values
$K=-r$, with $r=0,N,2N/3$, we must redefine the generators of the
$W(sl(N+3),sl(3))$ algebras (\ref{superWnew}) as follows:
\bea
W & = & \frac{1}{K+r}{\widetilde W}\quad ,\quad
G_a=\frac{1}{\sqrt{K+r}}{\widetilde G}_a \quad ,\quad
{\overline G}^a=\frac{1}{\sqrt{K+r}}{\widetilde{\overline G}}^a
\quad ,\quad r=0,N\quad ,
\label{redef0} \\
W & = & \frac{1}{(K+r)^2}{\widetilde W}\quad ,\quad
G_a=\frac{1}{K+r}{\widetilde G}_a \quad ,\quad
{\overline G}^a=\frac{1}{K+r}{\widetilde{\overline G}}^a
\quad ,\quad r=\frac{2N}{3} \quad .\quad
\label{redef} 
\eea
After the above redefinitions the OPEs (\ref{superWnew}) describe,
for $K=-r$, some uninteresting algebras obtained as a contraction of
$W(sl(N+3),sl(3))$. For the first two levels
$K=0,-N$ the central charge $c_u$ does not vanish, whereas
$c_u=0$ at $K=-2N/3$. In all three $K=-r$ cases one has
$c_g=\epsilon c_w=0$ and $c_t\neq 0$. Hence the algebra described by
the OPEs (\ref{superWnew}) exists for all values of the level parameter
$K$, except $K=-1-N$. However one has to keep in mind that the
algebras of interest to us correspond to $K+r\neq 0$.

Let us note that the spin-4 current $U_4(z)$ is defined to be primary 
with respect to the stress tensor $T_w (z)$\footnote{Due to
the regular OPE of the spin-3 current $W(z)$ with $U(z)$ and $J_a^b (z)$, 
$W(z)$ is a primary current, not only with respect to $T$,
but also with respect to $T_w$.}
\be
T_w (z_1)T_w (z_2) =   \frac{c_w}{2z_{12}^4}+
        \frac{2T_w}{z_{12}^2}+\frac{T_w'}{z_{12}} , \quad 
T_w (z_1)U_4(z_2)  =   \frac{4U_4}{z_{12}^2}+\frac{U_4'}{z_{12}} 
\ee
and both $T_w$ and $U_4$ have regular OPEs with the $u(1)$ and $sl(N)$
currents $U(z)$ and $J_a^b (z)$
\be
U(z_1)T_w (z_2) = U(z_1)U_4(z_2) = 
J_a^b (z_1)T_w (z_2) = J_a^b (z_1)U_4(z_2) = 
\mbox{regular}.  \label{coset}
\ee
Thus, it follows from the above considerations that
the currents $T_w (z),W(z)$ and $U_4(z)$ belong to the
coset $W((sl(N+3),sl(3))/(u(1)\oplus sl(N))$.

\setcounter{equation}0
\section{$W_3$ contractions of $W(sl(N+3),sl(3))$ algebras}

In this Section we would like to briefly recall the method for
constructing the so called realizations modulo null fields
for the $W_3$ algebra \cite{a6,a7,a8,a9,a10,a11,a12}. 
The distinguished feature of these
realizations, with respect to the ordinary ones, consists in 
allowing the presence
(besides standard terms) of a nonvanishing spin-4
{\it null} operator ${\cal V}$
in the OPE of the spin-3 current ${\cal W}$ with itself
\footnote{The spin-4 operator ${\cal V}$ could be composite or elementary.}
\bea
{\cal W}(z_1){\cal W}(z_2) & = & standard \; terms + 
\frac{{\cal V}}{z_{12}^2} +
\frac{{\cal V}'}{2z_{12}}\quad .  \label{nullW_3}
\eea
The {\it null} operator ${\cal V}$ must satisfy the following
requirement:
\bea
< {\cal V}{\cal V} >=0 \quad . \label{nullcond}
\eea
This corresponds to requiring that the OPE of
the operator ${\cal V}$ with itself contains no central term.
Obviously, being a null operator, ${\cal V}$ can generate only null fields
in its OPEs. All such null operators span an ideal and can be
consistently set
to zero, yielding a $W_3$ algebra which closes only modulo null currents.
Thus, on the shell defined by these constraints, the spin-2 and spin-3
currents form a $W_3$ algebra.

From the above definition it is clear that the problem of the
construction of realizations modulo null fields is a
very complicated one. However, it can be
reduced to the easier task of constructing ordinary realizations, but
for a larger ${\cal W}$-algebra, containing more currents than $W_3$
and including 
the OPE \p{nullW_3} within the full set of its OPEs. Then, for some discrete 
values of
the central charge, the null field condition \p{nullcond} for the spin-4
operator of such ${\cal W}$-algebra
could be satisfied. Hence, for these specific
values of the central charge, the realizations of such larger
${\cal W}$-algebra form
simultaneously a realization of $W_3$ modulo null fields.

The $W(sl(N+3),sl(3))$
algebras constructed in the previous Section provides a class
of such kind of ${\cal W}$-algebras, larger than $W_3$ and including in their
OPEs a spin-4 current ${\cal V}$, according to \p{nullW_3}.
Having found the candidate algebras, all we need, in order to use
the abovementioned approach for the construction of realizations of
$W_3$ modulo null fields, is to determine the truncation condition
of these class of ${\cal W}$-algebras to $W_3$, i.e. find the spectrum
$c_{w}$ corresponding to solutions of \p{nullcond}. As a second step,
we must build ${\cal W}$-algebras realizations for the specific
$c_w$ values, for which ${\cal W}$ reduces to $W_3$. We start with
nonlinear $W(sl(N+3),sl(3))$ algebras because they possess the following
properties:
\begin{itemize}
\item the algebras $W(sl(N+3),sl(3))$ include the OPE \p{nullW_3};
\item the algebras $W(sl(N+3),sl(3))$ can be linearized (see Section 4
below).
\end{itemize}
These facts are of use, because they respectively allow the following:
\begin{itemize}
\item to truncate (contract) the algebras $W(sl(N+3),sl(3))$ to $W_3$;
\item to build realizations, starting from the linearizing
$W(sl(N+3),sl(3))$ algebras. 
\end{itemize}

Indeed, inspecting (\ref{superWnew}) it is easy to 
realize that the OPE of the spin-3
current with itself looks like \p{nullW_3}.
We stress that it is not possible to perform a redefinition of
the currents of $W(sl(N+3),sl(3))$, such as to avoid the appearance of
the $U_4(z)$ current in the r.h.s. of the
OPE $W(z_1)W(z_2)$. Therefore, the
$W(sl(N+3),sl(3))$ algebras do not contain $W_3$ as a subalgebra.

The vacuum expectation values for 
the currents $T_w , W, U_4$ which are contained in
the coset $W((sl(N+3),sl(3))/(u(1)\oplus sl(N))$ read
\bea
< T_w T_w > & = &  -\frac{(2K+N)(-1+3K+2N)(1+4K+3N)}{2(K+N)(K+N+1)}
\quad , \nn \\
< WW > & = & 
-\frac{(-2+3K+2N)(2+5K+4N)}{27K(3K+2N)}
<T_w T_w > 
\quad , \nn \\
< U_4U_4 > & = & 
\frac{8N(-1+2K+N)(-3+3K+2N)(1+3K+2N)(4K+3N)(3+6K+5N)}
{120K^3-32K-32K^2-27N-59KN+230K^2N-27N^2+145KN^2+30N^3}
\nn \\
    & &
\times\frac{1}{3K^2(3K+2N)} <WW > \quad . 
\eea
The spin-4 operator $U_4(z)$
becomes a null field - under the condition $<U_4U_4>=0$, but
requiring at the same time that $W(z)$ and $T_w(z)$ themselves
do not turn into null fields, i.e. $<WW>\neq 0$ and $<T_wT_w>\neq 0$ -
at the following values of the level parameter $K$
and for the corresponding values of the
central charge $c_{w}$ :
\bea
K & = & \frac{1-N}{2}, (N > 1)\quad ;\quad
K=\frac{-1-2N}{3}\quad ,\quad
K=\frac{-3N}{4}  \quad
\Rightarrow           \quad
c_{w}=-2       \quad ,
  \label{spectrum0} \\
K & = & \frac{3-2N}{3}\quad
\Rightarrow    \quad
c_{w}=\frac{2(N-6)(N+15)}{(N+3)(N+6)} 
\quad , \label{spectrum1} \\
K & = & \frac{-3-5N}{6},(N\neq 3)\quad
\Rightarrow                                \quad
c_{w}=\frac{2(N+5)(2N+3)}{N-3} 
\quad . \label{spectrum}
\eea
Hence, just for these values of the level
parameter $K$, every realization of
the algebras \p{superWnew} induces a corresponding
realization - modulo null fields - of the $W_3$ algebra formed by the 
currents $T_w$ and $W$.
All other poles
and zeros of the vacuum expectation value $<U_4U_4>$ provide us with
further contractions of the algebra, where the
spin-3 current $W(z)$ and even the stress tensor 
$T_w$ become null operators.

A list of the central charges of the spectrum \p{spectrum1}
corresponding to some first  values of $N$ reads as follows:
\be
c_w=-\frac{40}{7},-\frac{17}{5},-2,-\frac{38}{35},
-\frac{5}{11},0,\frac{22}{65},\frac{46}{77},\frac{4}{5},\frac{25}{36}\quad .
\ee
Analogously, we can list the central charges of the spectrum \p{spectrum}
corresponding to first values of $N$ as follows:
\be
c_w=-30,-98,198,130,110,
102,\frac{494}{5},98,\frac{690}{7}\quad .
\ee
A first attempt to classify the possible algebras which allow a contraction
to $W_N$ is made in \cite{a12} (see also references therein) where the
central charge spectrum \p{spectrum1} was conjectured.
However, the spectrum of central charges for the contraction of 
the $W(sl(N+3),sl(3))$ algebras to $W_3$ proposed in \cite{a12} is
not exhaustive. In the present work the central charge spectrum \p{spectrum}
and the spectrum of $K$-values corresponding to the point $c_w = -2$
of \p{spectrum0} are constructed for the first time.
A remark is in order concerning the latter point.
Indeed this point of the resulting spectrum is particularly interesting,
since it corresponds to three distinct representations of $W_3$
with arbitrary values of $N$.
Namely, it turns out according to our result \p{spectrum0} that
there exist three different null-fields realizations of $W_3$,
which have the same value of the central charge $c_w=-2$.

The natural continuation and completion of our reasoning brings up the 
task of constructing explicitly the realizations of the
$W(sl(N+3),sl(3))$ algebras, which in turn yield
realizations of $W_3$ modulo null fields. The tool of the conformal 
linearization method of Refs. \cite{a4,a5,ks1,a13}
provides a direct way to construct such null-fields realizations,
a fact that has been exploited for the first time in Ref. \cite{a6}.
\setcounter{equation}0
\section{The linearizing $W(sl(N+3),sl(3))$ algebras}

The method of the conformal linearization \cite{a4,a5,ks1,a13}
is a tool that drastically simplifies the problem
of constructing explicitly the realizations - modulo null fields - of the
$W_3$ algebra obtained by contracting the
$W(sl(N+3),sl(3))$ algebras, according to the central charge spectrum
determined in the previous section. This is so, because any
realization of the linearizing
$W(sl(N+3),sl(3))$ algebras gives rise to a realization of the corresponding
nonlinear
$W(sl(N+3),sl(3))$ algebras.

The linearizing algebras for $W(sl(N+3),sl(3))$ can be constructed
starting with the currents
$({\cal T},{\cal U},{\cal J}_a^b,{\cal U}_1,{\cal G}_a$,
${\overline {\cal G}}^a,{\overline {\cal Q}}^a,{\cal W})$.
The OPEs of the conformally linearized
$W(sl(N+3),sl(3))$ algebras read
\begin{eqnarray}
{\cal T}(z_1){\cal T}(z_2) & = &  \frac{c_l}{2z_{12}^4}+
   \frac{2{\cal T}}{z_{12}^2}+\frac{{\cal T}'}{z_{12}}\quad ,
\quad 
{\cal T}(z_1){\cal J}_a^b (z_2)  =  \frac{{\cal J}_a^b}{z_{12}^2}+
\frac{{\cal J}_a^b{}'}{z_{12}} \quad ,\nn\\
{\cal T}(z_1){\cal G}_a(z_2) & = &  \frac{{\cal G}_a}{z_{12}^2}+
\frac{{\cal G}_a{}'}{z_{12}}\quad , \quad
{\cal T}(z_1){\overline {\cal G}}^a(z_2)  =   
\frac{{\overline {\cal G}}^a}{z_{12}^2}+
\frac{{\overline {\cal G}}^a{}'}{z_{12}} \quad ,\nn \\ 
{\cal T}(z_1){\overline {\cal Q}}^a(z_2) & = &  
\frac{3(K+1+N)+3+N}{2(K+1+N)}
\frac{{\overline {\cal Q}}^a}{z_{12}^2}+
\frac{{\overline {\cal Q}}^a{}'}{z_{12}} \quad ,\quad
{\cal T}(z_1){\cal U}(z_2)  =   \frac{{\cal U}}{z_{12}^2}+
\frac{{\cal U}'}{z_{12}} \quad ,\nn  \\
{\cal T}(z_1){\cal W}(z_2) & = &  
\frac{3(K+1+N)+3+N}{2(K+1+N)}
\frac{{\cal W}}{z_{12}^2}+\frac{{\cal W}'}{z_{12}} \quad ,\quad
{\cal T}(z_1){\cal U}_1(z_2)  =   \frac{{\cal U}_1}{z_{12}^2}+
\frac{{\cal U}_1{}'}{z_{12}} \quad ,\nn  \\
{\cal U}(z_1){\cal U}(z_2) & = &  \frac{2(N+1)(K+1+N)}{(3+N)z_{12}^2} 
\quad ,  \quad
{\cal U}(z_1){\overline {\cal Q}}^a(z_2)  =   -
\frac{{\overline {\cal Q}}^a}{z_{12}} \quad ,  \nn  \\  
{\cal U}(z_1){\cal W}(z_2) & = &  -
\frac{{\cal W}}{z_{12}} \quad ,  \quad
{\cal J}_a^b (z_1){\overline {\cal Q}}^c(z_2)  =   -(\delta_a^c 
{\overline {\cal Q}}^b
+\frac{1}{N}\delta_a^b {\overline {\cal Q}}^c )\frac{1}{z_{12}} 
\quad ,  \nn  \\
{\cal J}_a^b (z_1){\cal J}_c^d(z_2) & = & K(\delta_a^d\delta_c^b -
\frac{1}{N}
\delta_a^b\delta_c^d )\frac{1}{z_{12}^2}+(\delta_c^b {\cal J}_a^d -
\delta_a^d {\cal J}_c^b )\frac{1}{z_{12}}\quad ,\nn\\
{\cal J}_a^b (z_1){\cal G}_c(z_2) & = &  (\delta_c^b {\cal G}_a -\frac{1}{N}
\delta_a^b {\cal G}_c )\frac{1}{z_{12}} \quad ,  \quad
{\cal J}_a^b (z_1){\overline {\cal G}}^c(z_2)  =   -(\delta_a^c 
{\overline {\cal G}}^b
+\frac{1}{N}\delta_a^b {\overline {\cal G}}^c )\frac{1}{z_{12}} 
\quad ,  \nn  \\
{\cal U}_1(z_1){\cal U}_1(z_2) & = &  \frac{KN}{(1+N)z_{12}^2} 
\quad ,  \quad
{\cal U}_1(z_1){\overline {\cal Q}}^a(z_2)  =   
\frac{{\overline {\cal Q}}^a}{(N+1)z_{12}} \quad ,\nn  \\
{\cal U}_1(z_1){\cal G}_a(z_2) & = &  -\frac{{\cal G}_a}{z_{12}} 
\quad ,  \quad
{\cal U}_1(z_1){\overline {\cal G}}^a(z_2)  =   
\frac{{\overline {\cal G}}^a}{z_{12}} \quad ,\nn  \\
{\cal U}_1(z_1){\cal W}(z_2) & = &  -
\frac{N{\cal W}}{(N+1)z_{12}} \quad ,  \quad 
{\cal G}_a(z_1){\overline {\cal Q}}^b(z_2)  =   -
\frac{\delta_a^b {\cal W}}{z_{12}} \quad ,  \nn  \\
{\cal G}_a(z_1){\overline {\cal G}}^b(z_2) & = &  
\frac{K\delta_a^b}{z_{12}^2}+
\left[ {\cal J}_a^b -\frac{N+1}{N}\delta_a^b {\cal U}_1 \right]
\frac{1}{z_{12}} \quad ,  \quad
{\overline {\cal G}}^b(z_1){\cal W}(z_2)  =  -
\frac{{\overline {\cal Q}}^a}{z_{12}}  
\quad    .
\label{LinW}
\end{eqnarray}
In this basis the central charge of the
linearizing algebras
has the following expression:
\be
c_l  =  \frac{-6{(K+1+N)}^2+(N^2+2N+14)(K+1+N)+N^3+3N^2+2N+6}{K+1+N}
\quad    .
\label{Linc}
\ee 
The transformations that give the currents of the nonlinear
$W(sl(N+3),sl(3))$ algebras
in terms of the currents of the linearizing algebras \p{LinW} 
read\footnote{The expression for the $W$ current can be easily obtained from
the first pole of the OPE $G_a(z_1)\overline{G}^b(z_2)$ \p{superWnew}. 
It is a rather complicated and lengthy expression
and we do not write it here explicitly.}

\begin{eqnarray}
T & = & {\cal T}+\frac{(N+3)K{\cal U}{}'}{2(N+1)(K+1+N)}+{\cal U}_1{}'
\quad , \nn  \\
U & = & -\frac{2N{\cal U}}{N+1}+2{\cal U}_1
\quad , \quad
J_a^b  =  {\cal J}_a^b +\frac{1}{N}{\cal U}_1\delta_a^b
\quad , \quad
G_a  =  {\cal G}_a
\quad , \nn  \\
{\overline G}^a & = & {\overline {\cal Q}}^a
+h_1({\cal T}\;{\overline {\cal G}}^a)+
h_2({\cal G}_b\;{\overline {\cal G}}^a\;{\overline {\cal G}}^b)
+h_3({\cal U}_1\;{\cal U}_1\;{\overline {\cal G}}^a)+
h_4({\cal U}_1{}'\;{\overline {\cal G}}^a)
+h_5({\cal U}_1\;{\overline {\cal G}}^a{}')
\nn \\
    & &
+h_6({\cal U}\;{\cal U}_1\;{\overline {\cal G}}^a)
+h_7({\cal U}\;{\cal U}\;{\overline {\cal G}}^a)+
h_8({\cal U}\;{\overline {\cal G}}^a{}')+
h_9({\cal U}'\;{\overline {\cal G}}^a)+
h_{10}{\overline {\cal G}}^a{}''
+
\nn \\
    & &
+h_{11}({\cal J}_b^c\;{\cal J}_c^b\;{\overline {\cal G}}^a)
+h_{12}({\cal J}_b^c\;{\cal J}_c^a\;{\overline {\cal G}}^b)+
h_{13}({\cal U}\;{\cal J}_b^a\;{\overline {\cal G}}^b)
\nn \\
    & &
+h_{14}({\cal J}_b^a\;{\cal U}_1\;{\overline {\cal G}}^b)+
h_{15}({\cal J}_b^a{}'\;{\overline {\cal G}}^b)+
h_{16}({\cal J}_b^a\;{\overline {\cal G}}^b{}')
\quad . \label{TransLin} 
\end{eqnarray}
For the expressions of the coefficients $h_1$,...,$h_{16}$ see Table 2
given in appendix.

With the above expressions we have completed our task to construct
the realizations for
the $W(sl(N+3),sl(3))$ \p{superWnew} algebras, in terms of the currents of 
the linear
algebras (\ref{LinW}). In fact, starting from any given realization
of (\ref{LinW}), the use of (\ref{TransLin}) allows us to construct the
realizations of the nonlinear algebras (\ref{superWnew}). Notice the
appearence of the following exceptional value of the parameter $K$:
$K=0$, which is a singular point of the transformations
(\ref{TransLin}). Precisely for this value, the currents generating
the starting nonlinear
algebras (\ref{superWnew}) have to be redefined,
in order to prevent
the coefficients appearing in their OPEs from being singular, as it
appears from some of the expressions given in Table 1.

In addition, there is a second level $K=-1-N$ where the transformations
(\ref{TransLin}) have a singularity.
However, for this point the nonlinear algebras we started with,
i.e. (\ref{superWnew}), are not defined. Hence we must exclude this point
from our present consideration of the linearizing
$W(sl(N+3),sl(3))$ algebras as well. Thus, the set of
exceptional points where the realizations of the
$W(sl(N+3),sl(3))$ algebras develop some singular terms
is entirely determined by the structure relations of the
$W(sl(N+3),sl(3))$ algebras. Null field realizations of the
$W_3$ algebra that lies in
the coset $W((sl(N+3),sl(3))/(u(1)\oplus sl(N))$
are obtained from (\ref{TransLin}) at each value
of the parameter $K$ and  the corresponding central charge
$c_w$, given in eqs. (\ref{spectrum0})-(\ref{spectrum}).

\setcounter{equation}0
\section{Discussion and outlook}
In this paper we have explicitly constructed the algebras
$W(sl(N+3),sl(3))$. 
Using the relations between
nonlinear and linearizing algebras, we have found their
realizations, including the induced realizations of $W_3$ modulo
null fields, when the algebras
$W(sl(N+3),sl(3))$ are contracted to the $W_3$ one.
Such null field realizations exist for the following values of the
central charge of the $W_3$ algebra that lies in
the coset $W((sl(N+3),sl(3))/(u(1)\oplus sl(N))$:
\bea
c_{w}=-2,\quad 
\frac{2(N-6)(N+15)}{(N+3)(N+6)},\quad
\frac{2(N+5)(2N+3)}{N-3}. 
\eea
It is interesting to observe that one can rewrite this spectrum 
in terms of the spectrum of $W_3$ minimal models, as follows:
\bea
c_{w} = c_{W_3}^{min.mod.}(3,2),\quad c_{W_3}^{min.mod.}(5+N,2+N),\quad
c_{W_3}^{min.mod.}(4-N,6),
\eea
where
\begin{eqnarray}
c_{W_3}^{min.mod.}(p,q) & = & 2\Biggr[1-12\frac{(p-q)^2}{pq}\Biggr]
\quad  . \label{W3cm}         \end{eqnarray}

We wish to conclude this Letter by predicting the spectrum
of central charges for
the $W(sl(N+3),sl(3))$ minimal models.
One can start with noticing that the
conformally linearized
$W(sl(N+3),sl(3))$ algebras are homogeneous in
the currents 
${\overline {\cal Q}}^a,{\cal W}$. This remark implies that the latter
are null fields and can be consistently set to zero
${\overline {\cal Q}}^a={\cal W}=0$.
Inserting the above conditions into the expressions (\ref{TransLin})
leaves us with the Miura realization of
the nonlinear $W(sl(N+3),sl(3))$ algebras
in terms of the currents
${\cal T}$, ${\cal U}$, ${\cal J}_a^b$, ${\cal U}_1$, ${\cal G}_a$,
${\overline {\cal G}}^a$.
It is possible to introduce a decoupling basis in
the linearizing $W(sl(N+3),sl(3))$ algebras, with some redefined
energy-momentum tensor $T_{Vir}$,
which commutes with the remaining currents and possesses the
following central charge:
\begin{eqnarray}
c_{Vir} & = & 1-6\frac{(K-1)^2}{K}
\quad . \label{Virc}         \end{eqnarray}
The values of $c_{Vir}$ corresponding to the minimal models of the Virasoro
algebra at
\begin{eqnarray}
K=\frac{p}{q}
\Rightarrow c_{Vir}  =  1-6\frac{(p-q)^2}{pq}
\quad  \nn    \end{eqnarray}
induce the following spectrum for the central charge \p{ucoset} of
the nonlinear $W(sl(N+3),sl(3))$ algebras:
\be
c_{W(sl(N+3),sl(3))}^{min.mod.}  = 
N^2+26N+50-\frac{24p^2+(N+4)(N+3)(N+2)q^2}{pq}
\quad    .
\label{minc}
\ee 

In all known cases of linearizing algebras the minimal models
the Virasoro algebra spanned by $T_{Vir}$ reproduce the minimal
models for the corresponding nonlinear algebras. For instance,
in the case of $N=2$ superconformal and $W_3^{(2)}$ algebras,
it can be checked that the above procedure yields the spectrum of
minimal models \cite{a13}. The values of $c_{Vir} $ corresponding
to the minimal models of the Virasoro algebra (\ref{Virc}) are also
found to induce the central charges $c_{W_N}^{min.mod.}$ for the
$W_N$ algebras \cite{a13}. Using this argument, it seems reasonable to
expect that this feature persists for 
$W(sl(N+3),sl(3))$ algebras, as well. Nonetheless, it is needed
that our conjecture in eq. (\ref{minc}) for
the spectrum of central charges for the minimal models of the
$W(sl(N+3),sl(3))$ algebras be checked by standard methods.

\section*{Acknowledgments}
We gratefully acknowledge
the GRAAL group at the University of Rome II and Dr. C. Preitschopf
at the Humboldt University in Berlin for making
their computational facilities available to us.
S.B. wishes to thank BLTP at JINR and Prof. D.~Shirkov 
for very kind hospitality when this investigation was undertaken.
Special thanks go to Prof. L.~Bonora and Dr. R.~Percacci at SISSA for the 
hospitality that made possible to complete this work.
 
This investigation has been supported in part by the
Russian Foundation of Fundamental Research,
grant 96-02-17634, and INTAS, grant 94-2317 and grant of the
Dutch NWO organization.

\setcounter{equation}0
\section*{Appendix}

Next, we write down the expressions of the coefficients in the OPEs
(\ref{superWnew}) for the algebra $W(sl(N+3),sl(3))$.
$$
\begin{array}{|l|l|l|l|} \hline
 c_t = \frac{1}{R_1}\scriptstyle{(2R_1-24K^2-} &
 c_u = \frac{4NR_2}{3+N} &
 c_g =  -\frac{(2K+N)R_2}{3R_1} &
 \epsilon =  \frac{(2-R_2)(2R_1+R_2)}{9KR_2}\\
\scriptstyle{22KN-6N^2+KN^2)} &
 {\beta}_{1}  =  
 \frac{(R_2-2)(R_2-1)}{18K(K+N)R_2^2}\times &
 {\epsilon}_{1}  =  \frac{2N}{3K\epsilon} &
{\epsilon}_{14}  =  \frac{(3+N)(-3+K-N)}{144KNR_1\epsilon} \\
{\alpha}_1  =  \frac{(3+N)(2K+N)}{6NR_1} & 
  \scriptstyle{(R_1+R_2)(2R_1+R_2)} & 
  {\epsilon}_{2}  =  -\frac{2(3+N)}{3KR_2\epsilon} &
 {\epsilon}_{15}  =  
 \frac{(3+N)}{108KR_2\epsilon}\times \\ 
{\alpha}_{2} = -\frac{(2K+N)R_2}{3KR_1} &
{\beta}_{2}  =  
\frac{(R_2-2)(2R_1+R_2)(KR_2-R_1)}{18K(K+N)R_2^2} &
{\epsilon}_{3}  =  - \frac{4}{3K\epsilon} & 
 \frac{(6R_1+6K^2+7KN+2N^2)}{R_1}\\
{\alpha}_3  =  -\frac{(3+N)(2+N)}{24N^2R_1} &
{\beta}_{3} =  \frac{(2-R_2)(3+N)(2R_1+R_2)}{12KNR_2^2} &
{\epsilon}_{4}  = \frac{2R_1}{9KR_2\epsilon} &
{\epsilon}_{16}  =  
\frac{2(3+N)}{27K^2N(K+N)R_1\epsilon} \\
{\alpha}_4  =  -\frac{R_2}{3KR_1} &
{\beta}_{4} =  \frac{(R_2-2)(2R_1+R_2)}{6K(K+N)R_2} &
{\epsilon}_{5} =  \frac{(2-K)(3+N)}{18KNR_2^2\epsilon} &
{\epsilon}_{17} =  
\frac{(3+N)(6K+2N+KN)}{18K^2NR_1R_2\epsilon} \\
{\alpha}_5  =  \frac{2-K}{6KR_1}    &
{\beta}_{5} =  -\frac{2R_1}{3KR_2} &
{\epsilon}_{6}  =  
\frac{2(2-K)}{9K^2R_2\epsilon} &
{\epsilon}_{18}  =  
\frac{(3+N)}{27K^2N(K+N)}\times\\
{\alpha}_{6}  =  -\frac{3+N}{6NR_2} &
{\beta}_{6}   = \frac{(3+N)(2NR_1+R_2)}{12KN^2R_2^2} &
{\epsilon}_{7}  =  -\frac{3+N}{9KR_2\epsilon} &
\frac{(R_2^2+3KR_2+3KNR_1)}{R_1R_2\epsilon}\\
{\alpha}_{7}  =  \frac{{(3+N)}^2}{144N^3R_1R_2^2}\times &
{\beta}_{7}  =  -\frac{3+N}{3KNR_2} &
{\epsilon}_{8} =  -\frac{3+N}{9KR_2\epsilon} &
{\epsilon}_{19}  =  -\frac{1}{9K^2(K+N)R_1\epsilon}  \\
\scriptstyle{(2NR_1+NR_2+2R_2)} &
{\beta}_{8}  =  -\frac{3+N}{12KN} &
{\epsilon}_{9}  =  -\frac{1}{12\epsilon} &
{\epsilon}_{20} =  
\frac{2-4R_1+K^2+KN}{18K^2(K+N)R_1R_2\epsilon} \\
{\alpha}_{8}  =  \frac{(3+N)}{36NR_1R_2}\times &
{\beta}_{9}   = \frac{(2-K)(3+N)(2+R_2)}{12KN{R_2}^2} &
{\epsilon}_{10}  =  
\frac{{(3+N)}^2}{864KN^3R_1{R_2}^3\epsilon}\scriptstyle{(3K^2N} & 
{\epsilon}_{21} =  -\frac{2K+N}{3K^2(K+N)R_1\epsilon}  \\
\scriptstyle{(3R_1+6K^2+7KN+2N^2)} &
{\beta}_{10}  =  \frac{1}{3K(K+N)} & 
\scriptstyle{-18K-24NR_1-2KN^2-4N^3)}& 
{\epsilon}_{22}  =  
\frac{3K^2R_1-12K^2+4N^2}{36K^2(K+N)R_1\epsilon} \\
{\alpha}_{9}  =  \frac{(K-2)(3+N)}{12KNR_1R_2} &
{\beta}_{11}  = \frac{R_1}{3K(K+N)R_2} &
{\epsilon}_{11}  =  
\frac{{(3+N)}^2}{72KN^2R_1R_2^2\epsilon}\times &
{\epsilon}_{23} =  
\frac{1}{36K^2(K+N)R_1R_2\epsilon}\times \\
{\alpha}_{10}  =  \frac{1}{9KR_1(K+N)} &
{\beta}_{12}  =  \frac{R_2}{6K(K+N)} &
\scriptstyle{(6K+6N+5KN+4N^2)}& 
\scriptstyle{(-20K-16K^2- 35K^3+9K^4-} \\
{\alpha}_{11}  = \frac{R_2-3K^2-3KN}{18KR_1(K+N)} &
{\beta}_{13}  = \frac{(K-2)(2+R_2)}{6K(K+N)R_2} &
{\epsilon}_{12} =  
\frac{(3+N)}{36K^2N^2R_1R_2^2\epsilon}\scriptstyle{(K^2N-} &
\scriptstyle{16N-32KN-78K^2N+15K^3N-}  \\
{\alpha}_{12}  =  \frac{3+N}{6KNR_1} &
{\beta}_{14}  =  \frac{1}{36K(K+N){R_2}^2}\scriptstyle{(4+8K+} &
    \scriptstyle{18K-8N-10KN-4N^2)} & 
\scriptstyle{16N^2-54KN^2+ 6K^2N^2-12N^3 )}\\
{\alpha}_{13}  =  -\frac{1}{6KR_1} &
\scriptstyle{13K^2+9K^3+18K^4 +8N+38KN+} & 
{\epsilon}_{13}  =  
\frac{3+N}{432KNR_1R_2^2\epsilon} 
\scriptstyle{(27K^3-}
& \\
{\alpha}_{14}  =  -\frac{2K+N}{3KR_1} &
\scriptstyle{39K^2N+33K^3N+20N^2+46KN^2+} & 
\scriptstyle{60-48K-105K^2-72N-150KN}& \\
{\alpha}_{15}  =  -\frac{R_1+2K^2+KN}{6KR_1} &
\scriptstyle{20N^2K^2+16N^3+4KN^3)} & 
\scriptstyle{+3K^2N-60N^2-38KN^2-16N^3)}& \\
\hline 
\end{array}
$$
\begin{center}{\bf Table 1}
\end{center}
Here we define $R_1=1+N+K$ and $R_2=3K+2N$.

Finally, we give the coefficients for the currents in the nonlinear basis
(\ref{TransLin}).
$$\begin{array}{|l|l|l|l|} \hline
h_{1} =  \frac{1}{3K}&
h_{2} =  -\frac{1}{3KR_1} &
h_{3}  =  - \frac{2+N+N^2}{6KN^2R_1} &
h_{4}  = \frac{-2+2K-N-N^2}{6KNR_1} \\
h_{5}  = \frac{1+2K+3N}{3KNR_1} &
h_{6}  =  \frac{3+N}{3KN(1+N)R_1}  &
h_{7}  =  -\frac{(2+N)(3+N)}{6K{(1+N)}^2R_1}&
h_{8}  =  -\frac{(3+N)(K+N)}{3K(1+N)R_1} \\
h_{9}  =  -\frac{(3+N)(K+N)}{6K(1+N)R_1}  &
h_{10} =  -\frac{R_1+2(K+N)^2}{6KR_1} &
h_{11} =  -\frac{1}{6KR_1} &
h_{12} =  -\frac{1}{3KR_1} \\
h_{13} =  -\frac{3+N}{3K(1+N)R_1} &
h_{14} = \frac{2}{3KNR_1} &
h_{15} = \frac{1-K}{3KR_1} &
h_{16} = \frac{1-2R_1}{3KR_1}  \\
\hline 
\end{array}
$$
\begin{center}{\bf Table 2}
\end{center}

\end{document}